# Reliable Synthesis of Large-Area Monolayer $WS_2$ Single Crystals, Films, and Heterostructures with Extraordinary Photoluminescence Induced by Water Intercalation

*Qianhui Zhang, Jianfeng Lu, Ziyu Wang, Zhigao Dai, Yupeng Zhang, Fuzhi Huang, Qiaoliang Bao, Wenhui Duan\*, Michael S. Fuhrer, Changxi Zheng\**

Q. Zhang, W. H. Duan, C. Zheng
Department of Civil Engineering
Monash University
Clayton, VIC 3800 Australia
E-mail: wenhui.duan@monash.edu
E-mail: changxi.zheng@monash.edu

J. Lu
School of Chemistry
Monash University
Clayton, VIC 3800 Australia

Z. Wang, Z. Dai, Y. Zhang, Q. Bao
Department of Material Science and Engineering
Monash University
Clayton, VIC 3800 Australia

F. Huang
State Key Lab of Advanced Technology for Materials Synthesis and Processing
Wuhan University of Technology
Wuhan, 430070, China

Q. Bao, W. H. Duan, M. S. Fuhrer, C. Zheng
Monash Centre for Atomically Thin Materials
Monash University
Clayton, VIC 3800 Australia

Q. Bao, M. S. Fuhrer
ARC Centre of Excellence in Future Low-Energy Electronics Technologies (FLEET)
Monash University
Clayton, VIC 3800 Australia

**Abstract**

Two-dimensional (2D) transition metal dichalcogenides (TMDs) hold great potential for future low-energy optoelectronics owing to their unique electronic, optical, and mechanical properties. Chemical vapor deposition (CVD) is the technique widely used for the synthesis of large-area TMDs. However, due to high sensitivity to the growth environment, reliable synthesis of monolayer TMDs via CVD remains challenging. Here we develop a controllable CVD process for large-area synthesis of monolayer $WS_2$ crystals, films, and in-plane graphene-$WS_2$ heterostructures by cleaning the reaction tube with hydrochloric acid, sulfuric acid and aqua regia. The concise cleaning process can remove the residual contaminates attached to the CVD reaction tube and crucibles, reducing the nucleation density but enhancing the diffusion length of $WS_2$ species. The photoluminescence (PL) mappings of a $WS_2$ single crystal and film reveal that the extraordinary PL around the edges of a triangular single crystal is induced by ambient water intercalation at the $WS_2$-sapphire interface. The extraordinary PL can be controlled by the choice of substrates with different wettabilities.

**1. Introduction**

Semiconducting two-dimensional (2D) transition metal dichalcogenides (TMDs) have emerged as the platform for exploring exotic physics and developing next-generation low-energy electronics. In particular, the group 6 TMDs ($MoS_2$, $WS_2$, $MoSe_2$, and $WSe_2$) possess direct bandgaps leading to strong light–matter interaction when reaching monolayer thickness.[1-4] Other reported distinguishing properties of monolayer TMDs include large spin-orbit coupling,[5] valley degeneration,[6] and high carrier mobility.[7-9] With their high flexibility, high Young's modulus,[10-12] and characteristic optical response to strain,[13-15] atomically thin TMDs are promising candidates for a novel class of flexible electronics. In particular, due to reduced Coulomb screening together with strong quantum confinement, strong excitonic binding energies have been observed in monolayer TMDs up to 1 eV,[16-18] leading to the observation of excitons, trions, and biexcitons at room temperature.[19,20] The recent demonstration of tuning excitons and trions using dielectric engineering also suggests an attractive way to modulate the optical properties of TMDs for future applications.[21, 22] Despite these broad potentials, growth of large-area TMD monolayers is undoubtedly the prerequisite.

So far, chemical vapor deposition (CVD) has been the widely used method enabling the fabrication of high-quality TMDs crystals and films of scalable size.[23] Intensive studies of the growth of monolayer $MoS_2$,[24-27] $WS_2$,[28-32] $MoSe_2$,[33-35] and $WSe_2$[36-38] in the TMD family have been conducted in recent years, with yields of crystals of various size and quality. Moreover, in-plane heterostructures of monolayer TMDs with graphene have been achieved recently on $MoS_2$[39-41] and $WS_2$[42]. Despite these exciting advances, challenges remain to the achievement of controllable synthesis. For example, due to high sensitivity to the growth condition, a critical issue for the CVD synthesis approach is repeatability. Also, many papers have reported nonuniform PL in single crystals of $WS_2$, yet similar optical inhomogeneity has not been observed in $WS_2$ thin films, and the inducing factors are still elusive.[43, 44]

In this paper, we report a reliable CVD synthesis method for obtaining large-area monolayer $WS_2$ single crystals (~ 200 µm) and films (~ 1 cm × 1 cm) of high quality and systematically compare the PL uniformity between single crystal and film. We found that the quality of CVD grown $WS_2$ was greatly affected by the cleanness of the CVD reaction quartz tube and crucibles. High-temperature oxygen gas ($O_2$) annealing was adopted to clean the tube and the crucibles before each batch of growth. With repeated usage, however, the residues accumulated in the CVD reaction chamber could not be effectively removed by this process. As a result, crystals grown using old tubes tended to display undesirable features such as small size and spots of multilayers, even with optimized parameters. Here we report an acid cleaning method for CVD growth quartz tube and crucibles that can effectively clean the residue adhering to the tube and crucibles. We found that by combining acid cleaning with high-temperature $O_2$ annealing of the quartz tube and crucibles after running multiple growth procedures, the CVD process could generate controllable and repeatable growth results of large-area monolayer crystals and films. The outcome was comparable to crystals grown by a new quartz tube and crucibles. Optical and morphological characterization of the grown $WS_2$ single crystals revealed that nonuniform PL emission along the edges of triangular single crystals was related to ambient water intercalation. Moreover, PL peak fitting indicated that the strong PL emission at the edges of single crystals was attributable to water intercalation to the interface between the $WS_2$ and the hydrophilic substrate. On the other hand, nonuniform PL behavior was not observed in large-area thin film or disappeared when the single crystals were transferred onto a hydrophobic substrate. Thus, by depositing the $WS_2$ single crystals on hydrophobic surfaces, controllable homogeneous PL emission can be achieved.

## 2. Results and Discussion

The monolayer $WS_2$ crystals/films were grown by CVD on atomically flat sapphire [$Al_2O_3(0001)$] substrates using precursors of $WO_3$ and S powders at elevated temperatures. The precursors and substrates loaded into separate quartz crucibles were located at upstream (S), middle ($WO_3$), and downstream (sapphire) of a 1 inch quartz tube, see **Figure 1**a. Prior to every usage, the crucibles and tube were cleaned by $O_2$ cleaning at 1050 °C for 1 h. During the growth, the temperatures of the $WO_3$ and S were raised to 860 °C and 180 °C, respectively, over 0.5 h, and monolayer $WS_2$ crystals/films were achieved by then maintaining the temperatures for 0.25 h. The growth was terminated by the fast cooling process. More details of the CVD process are given in the Experimental Section.

When crucibles and tubes that had never previously been used for any CVD growth were used, large-area $WS_2$ monolayers in the form of either triangular single crystal or continuous film could be synthesized with optimized parameters. Figure 1b,f show optical images of representative monolayer $WS_2$ single crystals with sizes up to 200 μm and continuous film covering almost the entire 1 cm × 1 cm substrate. The single crystals tended to grow on the substrate located close to the center of the furnace where the temperature and the diffusion velocity of $WS_2$ species were high whereas the vapor saturation and the nucleation rate were low. Meanwhile, $WS_2$ film tended to grow on the substrate further away where the temperature was lower but the crystal nucleation rate was higher.

The optical properties of the $WS_2$ crystals/film were investigated by scanning Raman/PL spectrum microscopy. Figure S1a,b present the Raman spectra taken from the $WS_2$ single crystal and the film, respectively. A strong $E'$ peak at 350 cm$^{-1}$ is observed. The corresponding PLs of the samples are given in Figure S1c and d, respectively. Using Lorentzian fitting, we could decompose both PLs into trion ($X^-$) and exciton ($X^0$) peaks. The formation of the negative trion peak was due to the n-type doping caused by S vacancies formed during the CVD growth.[45] The energy differences between trion and exciton peaks were 30 meV for the single crystal and 27 meV for the film, well within the earlier reported range of 20-40 meV of mechanically exfoliated monolayer $WS_2$.[46] Using Raman and PL for mapping, the optical uniformity of the single crystals could be identified, see Figure 1c-e. The region around the edge of the crystal shows much stronger PL intensity (Figure 1d), along with a redshift in the peak energy (Figure 1e), compared to the inner region, whereas the Raman intensity is quite uniform across the crystal (Figure 1c). The brightening is unlikely to be due to defect modulation at the crystal edge, because the width of the region is up to 4 μm rather than a line.[4] Moreover, atomic force microscopy (AFM) topography of the other crystal on the same substrate indicates a height difference between the outer and inner regions, see Figure S2b taken from a different crystal. We therefore inferred that the enhanced PL intensity around the crystal edge was caused by water intercalation. The detailed brightening mechanism is analyzed later together with the spectra shown in Figure 4. Due to the lack of water intercalation, the WS$_2$ film indicates uniform Raman, PL, and height across a large area in the inner region, see Figure 1g-i and Figure S2d.

Whereas the new quartz tube and crucibles could grow high-quality WS$_2$ crystals and films, the controllability of CVD growth and the quality of the crystals and films decreased continuously as the usage of the tube and crucibles increased. Figure 1j-l displays the CVD growth evolution of monolayer WS$_2$ crystals grown by the same quartz tube and crucibles using the same parameters as the number of growth rounds increases. It can be seen that the average size of the WS$_2$ crystals decreases gradually from ~200 μm to ~20 μm when the tube usage changes from new (Figure 1b, first batch of growth) to old (Figure 1l, 15th batch of growth). In particular, dark spots and multilayers appearing as white regions are observed in Figure 1l.

The cause of the growth degradation was the accumulated deposition of WS$_2$ particles on the wall of the quartz tube and the surfaces of the crucibles as their usage continued. During the CVD growth, WO$_3$ vapor was reduced by sulfur vapor with the assistance of H$_2$ to form the intermediate product WO$_{3-x}$, that was further reduced to WS$_2$. During the growth and the cooling stage, WS$_2$ deposited not only on the sapphire substrates, but also on the inner wall of the quartz tube and the surfaces of the crucibles. That residual deposition gradually altered the growth condition with increasing usage of the tube and crucibles. Although high-temperature O$_2$ annealing could remove the excessive deposition to some extent and increase the duration of repeatable growth, that cleaning approach was not effective due to the low evaporation rate of WO$_{3-x}$.

To further remove the residual contaminants, we developed an acid cleaning process for the used quartz tube and crucibles. As an example, in **Figure 2** the method is demonstrated on the quartz tube and crucibles after the 15th growth batch. First, the high-temperature O$_2$ annealing was carried out on the used tube and crucibles that were then sequentially cleaned by hydrochloric acid (HCl), sulfuric acid (H$_2$SO$_4$), and aqua regia [a mixture of nitric acid (HNO$_3$) and HCl in 1:3 molar ratio]. The crucibles were soaked in the acids contained in a large beaker, and the inner wall of the CVD reaction quartz tube was washed by the acids enclosed by sealing both ends of the tube, followed by swirling (Figure 2b). The acids help reduce the residual tungsten oxides transformed by the O$_2$ annealing and also detach the products from the surfaces of both tube and crucibles. The acid-cleaned tube and crucibles were then rinsed by water to wash off the residue of acids and the possible byproducts. This was followed by another high-temperature O$_2$ annealing before the next round of CVD growth, to further clean the tube and crucibles. The contrast between the quartz tube (crucibles) before and after acid cleaning is evident in Figure 2a,c (Figure S3). After acid cleaning and O$_2$ annealing, the quartz tube has a clear appearance.

With the same parameters, the growth of high-quality monolayer WS$_2$ undertaken using the acid-cleaned quartz tube and crucibles is retained, see Figure 2d. The maximum size of a WS$_2$ single crystal with uniform thickness could reach 176 µm. The overall quality is comparable to that obtained using the new tube and crucibles. The inset in Figure 2d shows a single crystal that is correlatively characterized by AFM, Raman, and PL (Figure 2e-g). Similarly, a band-like region along the crystal boundary presents much stronger PL than the inner region, whereas the Raman intensity across the crystal is quite uniform. The two regions showing differences in PL intensity indicate a height difference of ~ 0.37 nm that is the height of monolayer ice. The observations of height and PL intensity differences indicate the existence of water intercalation, which is consistent with the results of single crystals grown using a new tube and crucibles.

As well as the growth of monolayer WS$_2$ single crystal/film, the acid cleaning method could be applied to obtain reliable growth of large-area lateral graphene-WS$_2$ heterostructures. Following our previous work,[42] single-layer graphene was transferred to a sapphire substrate and patterned by photolithography and oxygen plasma etching. Because the surface energy of sapphire is higher than that of graphene, monolayer $WS_2$ tends to grow on the exposed sapphire surface and forms large-area films connecting to graphene edges under optimized growth parameters. Although growth of large-area graphene-$WS_2$ heterostructures can be achieved using new crucibles and tube, it is difficult to prepare uniform monolayer $WS_2$ films connecting to graphene after the tube and crucibles have been used for multiple growths, see Figure 3a,b. With the use of acid cleaning, however, the growth of large-area graphene-$WS_2$ heterostructures can again be obtained, see Figure 3a,c. The optical image of a large-area graphene-$WS_2$ heterojunction grown using an acid-cleaned tube and crucibles is shown in Figure 3d. The Raman intensity, PL intensity, and PL peak energy taken from the red square region of Figure 3d are given in Figure 3e-g, respectively. As shown, the overall PL intensity and peak energy of the region is uniform but the order of intensity is at a low level, comparable to that of the inner region of a single crystal and film. This demonstrates that there is no water intercalation under the heterojunction due to the lack of open edges.

In contrast to $WS_2$ film, that exhibited uniform PL characteristics, nonuniform PLs were observed on triangular $WS_2$ single crystals grown in all cases. As another example, **Figure 4**a presents the AFM topography of a small $WS_2$ crystal grown by the tube and crucibles after multiple usages without acid cleaning (denoted as 'old tube and crucibles' here). A 0.37 nm height increment caused by water intercalation is observed around the edge of the crystal, see the line profile presented in Figure 4a. The regions showing PL intensity enhancement (Figure 4c) and the redshift of PL peak energy (Figure 4d) are highly correlated with the location of water intercalation. PL inhomogeneity in $WS_2$ crystals has been observed by different groups due to edge boundaries, defects, chemical heterogeneity, and impurity doping.[4, 43, 44] Here we observe strong PL enhancement due to the water intercalation at the interface between sapphire and $WS_2$. This observation is unlike the results of other researchers. First, height differences at locations irradiating strong PL are clearly observed, whereas no height difference was observed on samples showing PL enhancement due to chemical heterogeneity and impurity doping.[43, 44] Secondly, no PL enhancement is observed in our WS$_2$ film and WS$_2$-graphene heterojunction. This result can eliminate the effect of chemical heterogeneity on the optical nonuniformity in our samples.

To understand the mechanism of PL enhancement caused by water intercalation, Figure 4e-g and e'-g' present the PLs of regions with water (Point A) and without water (Point B). The three crystals are the single crystals shown in Figure 1b, Figure 2d, and Figure 4a, respectively. It can be seen that the regions with intercalated water illuminate PL with 5-10 times higher peak intensity than the non-intercalated regions do when the regions are irradiated under the same laser power. However, all six PLs indicate a trion peak due to the intrinsic n-type doping caused by impurity and sulfur vacancies.[43, 44] Furthermore, biexciton emission at 1.930 eV and 1.952 eV is observed in water-intercalated regions of the two crystals grown by the new tube and the acid-cleaned tube (Figure 4f,g). Here we attribute the peak to biexcitons rather than to localized defects contributing to nonradiative recombination, because the region shows much stronger PL.[47] The appearance of biexciton emission suggests that the exciton density is high.

In the inner region of WS$_2$, the sapphire substrate is hydrophilic and terminated by Al.[48] The surface is a good electron acceptor. Thus the intrinsic n-type doping of WS$_2$ is depleted by the Al-terminated sapphire surface. However, when the ambient water molecules start intercalating to the interface between sapphire and WS$_2$ from the single crystal edges, monolayer ice indicating ~ 0.37 nm is formed by the water molecules in that confined region. This layer of ice is a very good dielectric and can effectively screen the influence of the sapphire substrate, releasing the intrinsic n-type doping of WS$_2$. Meanwhile, some intercalated water molecules can interact with sulfur vacancies to reduce the defect state within the bandgap to decrease the defect-mediated nonradiative recombination and increase the exciton lifetime.[47, 49] As a result, we observe stronger PL intensity. In comparison with the PLs taken from the regions without water intercalation, the exciton and trion peaks from water-intercalated region indicate redshifts due to the released n-type doping,[50] see Supplementary Table 1 for details of the exciton, trion, and biexciton energy.

Finally, to prove that the characteristic PL in the $WS_2$ single crystals was caused by water intercalation, we transferred CVD grown $WS_2$ crystals from the hydrophilic sapphire substrate to hydrophobic polystyrene (PS) via a polymer-assisted wet transfer method.[51] In brief, a thin layer of PS was spin-coated on the sapphire substrate with $WS_2$ crystals and then the substrate was immersed in the KOH solution, see **Figure 5**a. The immersed hydrophobic PS layer together with $WS_2$ crystals was immediately delaminated from the sapphire substrate due to water repulsion and floated on top of the KOH solution. Then the PS layer, together with the $WS_2$ crystals, was transferred to distilled (DI) water to rinse off the alkali residues. Afterwards, $SiO_2$/Si was applied to fish up the PS layer and form the layered structure of $WS_2$/PS/$SiO_2$/Si. Some details of the process are given in the Experimental Section. The optical image, PL intensity mapping and PL peak position mapping of a transferred crystal are presented in Figure 5b-d, respectively. As shown, no enhanced intensity is observed around the boundaries of the crystal. Moreover, the peak energy is quite uniform across the crystal, owing to the absence of water intercalation along the interface between the $WS_2$ crystals and the hydrophobic PS substrate. Figure 5e compares the PL spectra taken from the inner region of a $WS_2$ crystal on sapphire and the blue point shown in Figure 5c,d. Both PLs indicate similar features except for intensity due to the different transparency of the substrates. These results indicate that the uniformity of the PL emission across the $WS_2$ single crystal can be controlled by the choice of hydrophobic substrates.

## 3. Conclusion

In summary, our results show that the condition of the CVD environment plays a crucial role in WS$_2$ synthesis. The WS$_2$ growth suffers degradation as the usage of the CVD reaction tube and crucibles increases, due to the accumulation of excessive WS$_2$ particles on their walls and surfaces. A process of acid cleaning with HCl, H$_2$SO$_4$, and aqua regia provides a feasible solution to the problem, with efficient removal of the contaminants. By applying the acid cleaning process and high-temperature O$_2$ annealing, the synthesis of high-quality WS$_2$ single crystals, films, and in-plane heterostructures with graphene can be achieved in a controllable and repeatable manner. The PL and AFM characterizations of three WS$_2$ single crystals reveal that the intercalated water along the interface between WS$_2$ crystal and hydrophilic sapphire substrate plays an important role with regard to the PL characteristics. We infer that, for WS$_2$ single crystals, intercalated water can form monolayer ice to screen the electrostatic influence of the sapphire substrate and interact with the sulfur vacancy to reduce defect states, leading to the strong enhancement of PL intensity. The uniform PL emission is achieved by the transfer of WS$_2$ single crystals onto the hydrophobic PS surface, which further proves our proposed mechanism and indicates that the homogeneity of the PL emission is controllable by depositing the crystals onto hydrophobic substrates.

## 4. Experimental Section

*CVD growth*: Monolayer WS$_2$ single crystals, films, and heterostructures with graphene were grown by CVD using the setup schematically shown in Figure 1a. The precursor of WO$_3$ was positioned in the center of the CVD furnace. The sulfur precursor was placed in the upstream direction of the CVD furnace and heated separately by a heating tape. The WS$_2$ growth substrates were Al$_2$O$_3$(0001), located in a crucible immediately next to the WO$_3$ powder downstream. The substrates were cleaned by acetone and 2-propanol (IPA) ultrasonication for 10 min each before growth. During growth, a combination of ultra-high-purity hydrogen (5 sccm) and argon (200 sccm) was streamed at atmospheric pressure. The CVD furnace and the sulfur heating tape temperatures rose to 860 °C and 180 °C respectively in 0.5 h at the same rate and were maintained for 0.25 h. At elevated temperatures, the sulfur vapor was carried to the high-temperature region of the quartz tube to react with $WO_3$ vapor. This step was followed by a two-step fast cooling stage in which the furnace was first slightly then fully opened to allow the temperature to drop towards room temperature. Before each growth batch, the quartz tube and crucibles were cleaned by $O_2$ (50 sccm) at 1050 °C for 1 h before cooling naturally to room temperature.

*Acid cleaning of the tube and crucibles*: The CVD quartz tube and crucibles were first processed in an $O_2$ annealing session at 1050 °C for 1 h. After annealing, the tube was cleaned by HCl (mass fraction 32%), $H_2SO_4$ (mass fraction 98%), and aqua regia ($HNO_3$ and HCl with a molar ratio of 1:3) in turn. For each acid cleaning, both openings of the tube were sealed by rubber, with the acid contained in the tube. The tube was swirled manually to allow the acid to contact all regions of the inner surface and to help the acid react with the contaminants. The crucibles were placed in a large beaker containing the abovementioned acids in the same order and soaked for 0.5 h each to ensure reaction of the acids and the contaminants.

*Optical and AFM measurements*: Raman/PL measurements were performed by a confocal microscope system (WITec alpha 300R) with a 50× objective lens and a 532 nm laser excitation. A 600 line $mm^{-1}$ grating were used to collect the spectra. During PL and Raman mapping, 50 μW laser power and 1 s integration time were used for the prevention of sample damage. The PL intensity images were obtained via the summation of the PL intensity from 1.9 to 2.1 eV. The PL peaks of excitons, trions and biexcitons were estimated from Lorentzian fitting. The AFM measurement was carried out on a Bruker Dimension Icon in tapping mode.

*Polymer-assisted wet transfer*: Atomically thin $WS_2$ crystals on the as-grown sapphire substrate were spin-coated with PS ($M_w$~192000) in toluene solution (50 mg/mL) at 3000 rpm for 60 s and soft baked at 80 °C for 3 min. Next, edges of the spin-coated PS thin film on sapphire were scratched by a scalpel. Then, the PS film with $WS_2$ crystals attached was delaminated at the air surface of a 2 mol/L KOH solution at room temperature. The detached PS film was rinsed by DI water to remove the alkali residues. A rigid substrate such as $SiO_2$/Si was adopted to fish out the PS film from the top surface, leaving the $WS_2$ crystals sitting on top of the PS film supported by the rigid substrate.

**Acknowledgements**

C.Z. acknowledges support from ARC DECRA (DE140101555). M.S.F. acknowledges support from ARC (DP150103837 and FL120100038). This work was performed in part at the Melbourne Centre for Nanofabrication (MCN) in the Victorian Node of the Australian National Fabrication Facility (ANFF).

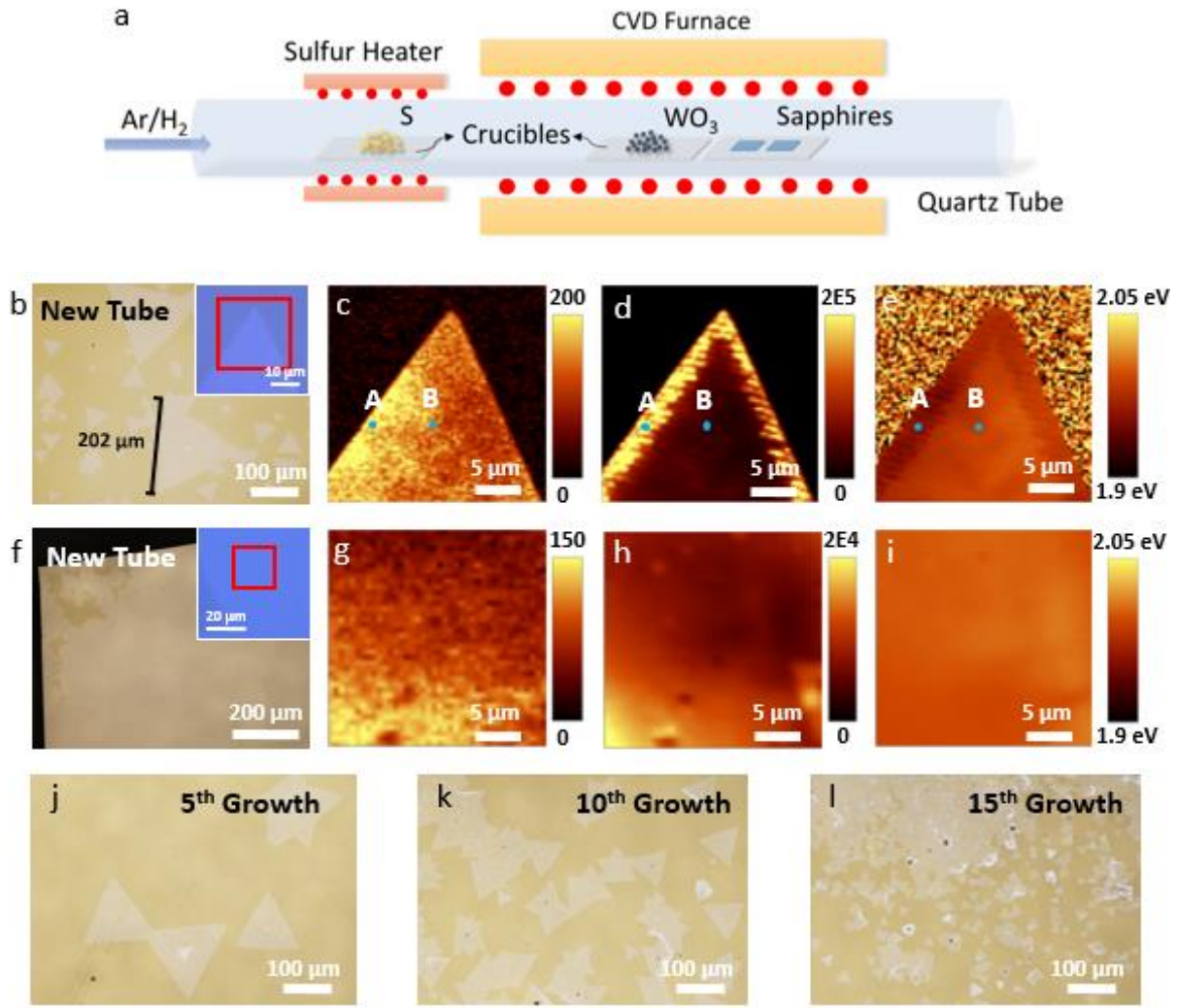

**Figure 1.** (a) Schematic of the CVD setup for $WS_2$ growth. (b) Optical image of monolayer $WS_2$ crystals on sapphire grown by a new CVD tube. Inset: zoomed-in image of a region of an as-grown $WS_2$ single crystal. (c-e) Raman intensity, PL intensity, and PL peak position mappings of the red square region in the inset of (b). (f) Optical image of monolayer $WS_2$ film on sapphire grown by a new CVD tube. Inset: zoomed-in image of a region of the film. (g-i) Raman intensity, PL intensity, and PL peak position mappings of the red square region in the inset of (f). (j-l) Optical images of as-grown $WS_2$ on sapphire by continuously used CVD tube and crucibles.

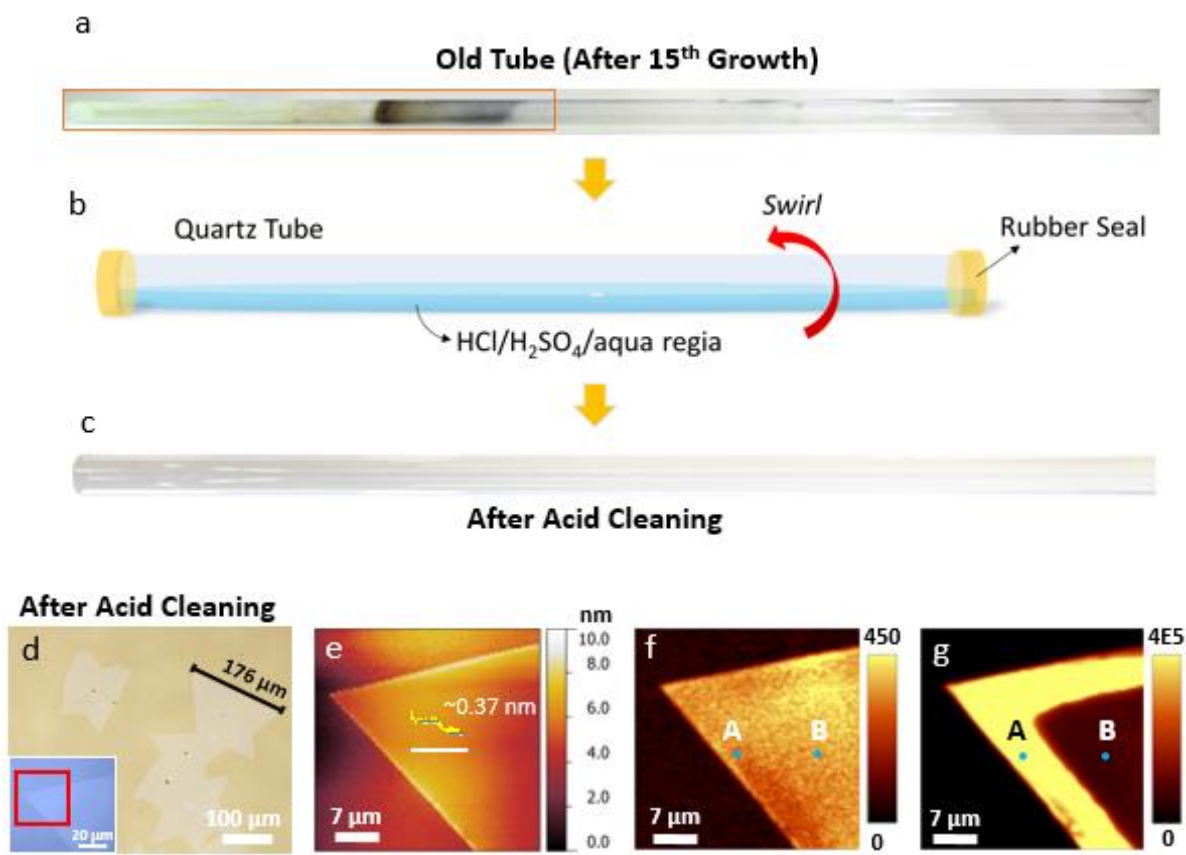

**Figure 2.** (a,c) Photograph showing the contrast between an old tube (after 15$^{th}$ growth) and the same tube after acid cleaning. (b) Schematic of the acid cleaning process for the CVD reaction tube. (d) Optical image of monolayer WS$_2$ crystals on sapphire grown by the acid-cleaned CVD tube. Inset: zoomed-in image of a region of an as-grown WS$_2$ single crystal. (e-g) AFM morphology, Raman intensity, and PL intensity mappings of the red square region in the inset of (d).

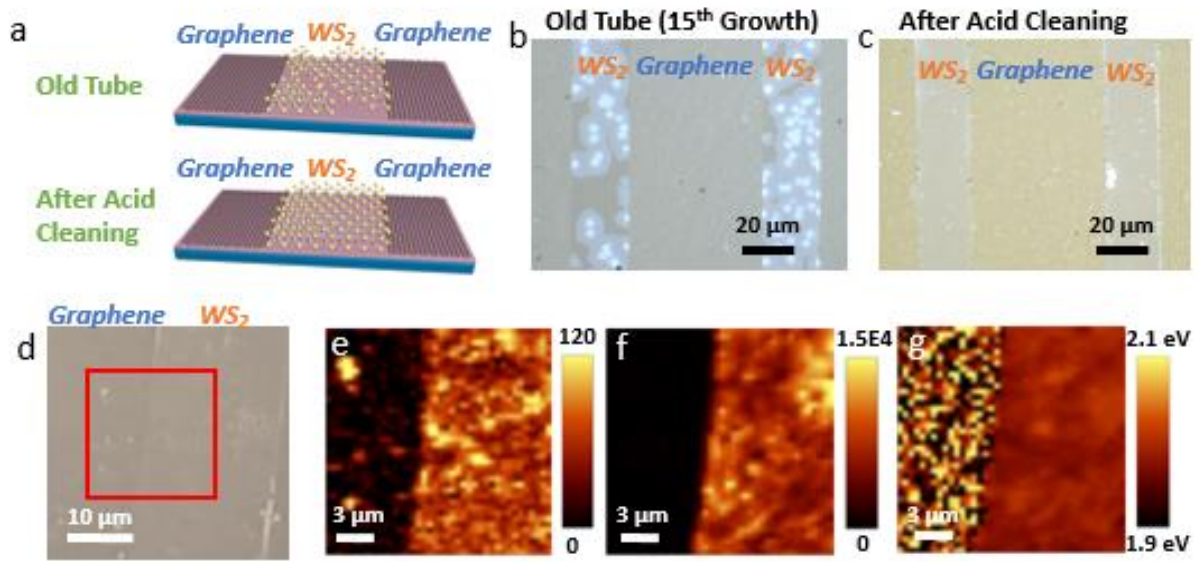

**Figure 3.** (a) Schematic of CVD grown in-plane graphene-WS$_2$ heterostructures using old tube and acid-cleaned tube. (b,c) Optical image of in-plane graphene-WS$_2$ heterostructures grown using an old tube and an acid-cleaned tube, respectively. (d) Zoomed-in optical image of a region of the as-grown graphene-WS$_2$ heterostructures. (e-g) Raman intensity, PL intensity, and PL peak position mappings for WS$_2$ film in the heterostructure of the red square region in (d).

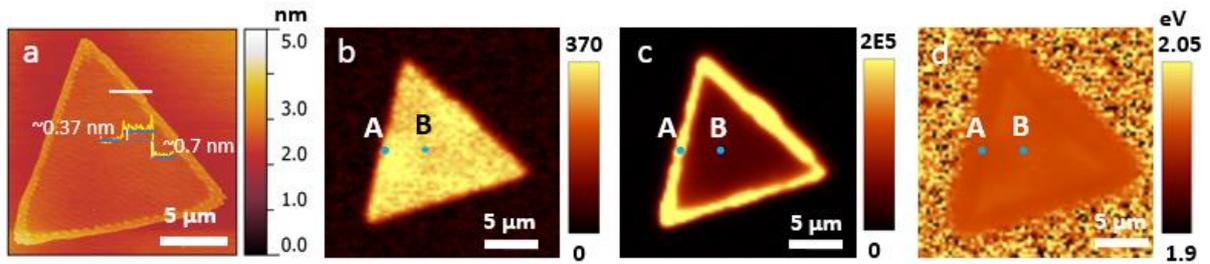

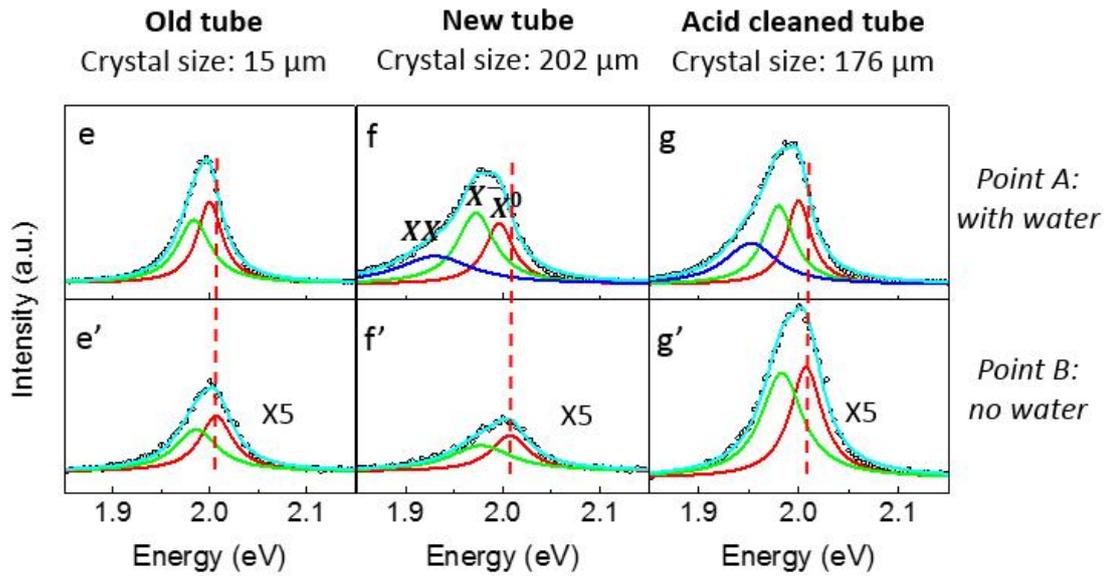

**Figure 4.** (a-d) AFM height, Raman intensity, PL intensity, and PL peak position mappings respectively for a small WS$_2$ single crystal grown by a multiply used CVD tube. (e-g, e'-g') PL spectra with Lorentz fitting at edge and inner regions for the small WS$_2$ crystal grown by the old tube and the two large WS$_2$ crystals grown by the new tube and the acid-cleaned tube.

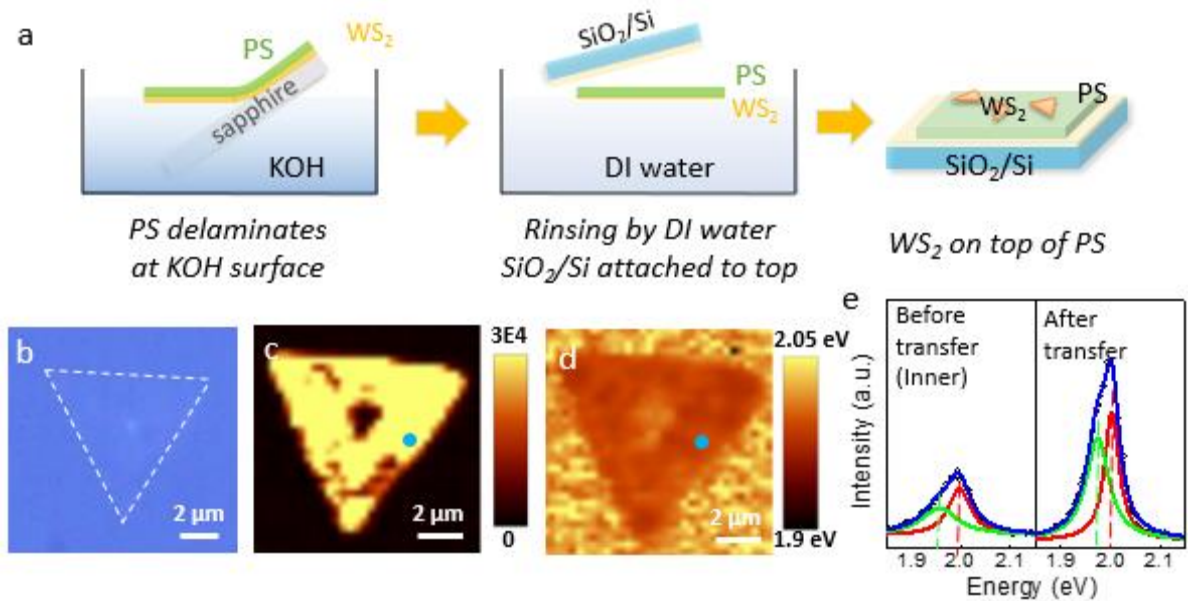

**Figure 5**. (a) Schematics of PS transfer process. (b-d) Optical image, PL intensity and peak position mappings of an as-transferred $WS_2$ crystal on PS. (e) PL spectra of the $WS_2$ crystal before and after transfer

Supporting Information

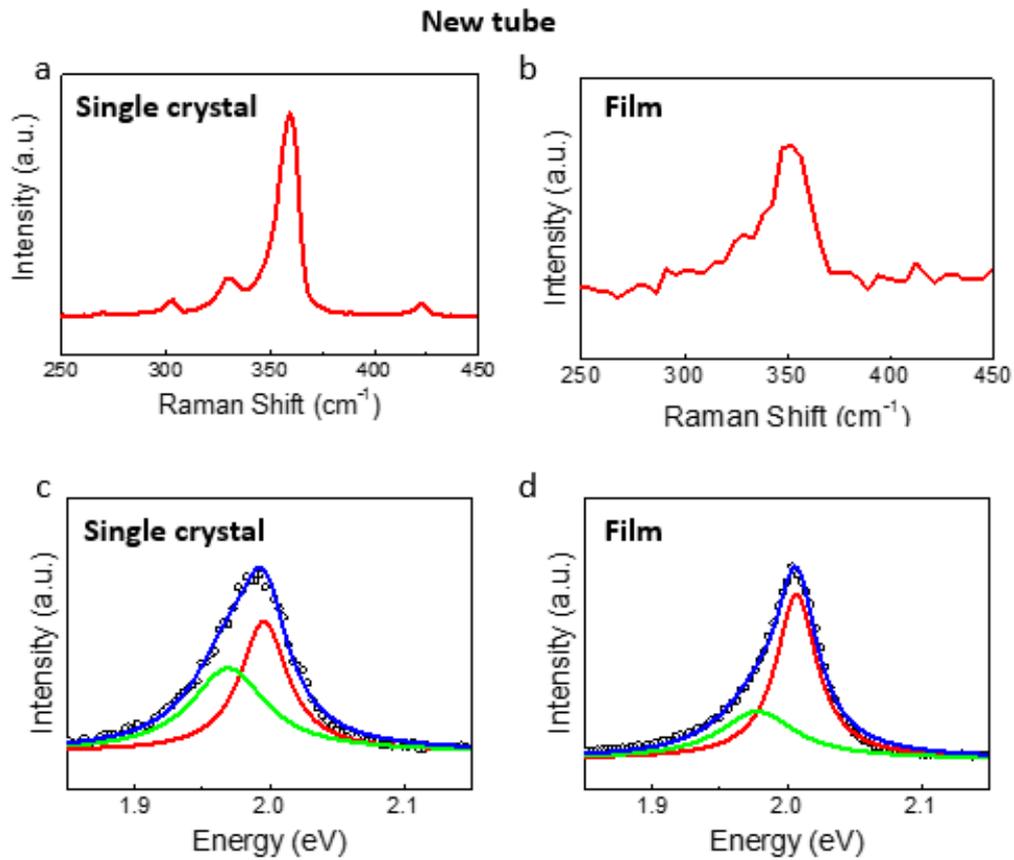

**Figure S1.** Raman and PL spectra of the CVD grown WS$_2$ using a new reaction tube. (a,b) Raman spectra collected from an as-grown WS$_2$ single crystal and film. (c,d) PL spectra collected from an as-grown WS$_2$ single crystal and film. Lorentzian fitting in (c) and (d) show the PL peak in each case is consist of a trion peak (green curve) and an exciton peak (red curve).

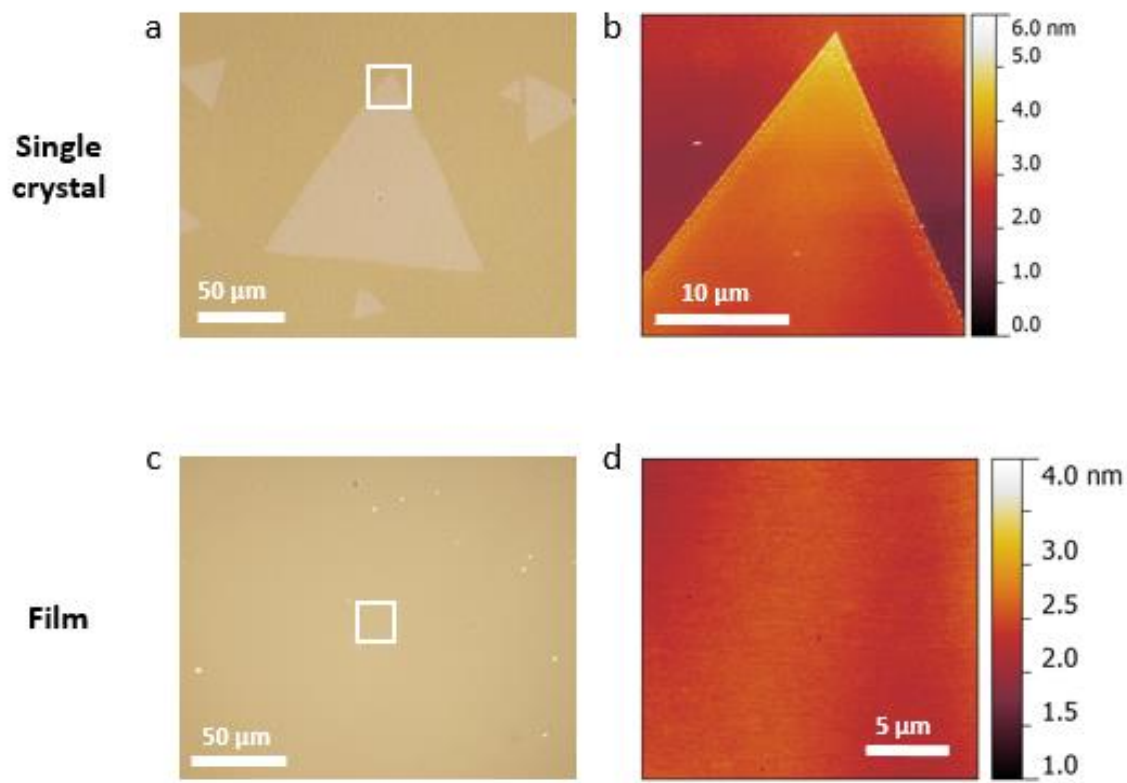

**Figure S2.** AFM characterization of the CVD grown $WS_2$ using a new reaction tube. (a) Optical image of a large as-grown $WS_2$ single crystal. (b) AFM height mapping of the region in the white square in (a). (c) Optical image of large-area as-grown $WS_2$ film. (d) AFM height mapping of the region in the white square in (c).

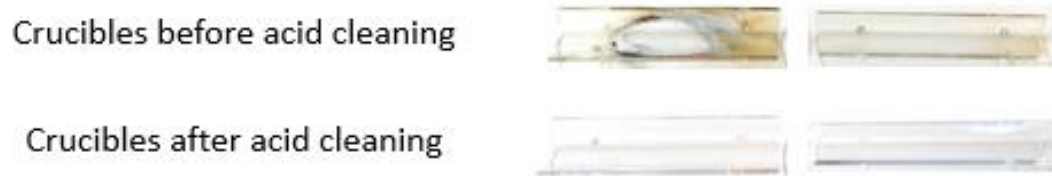

**Figure S3.** Photographs showing contrast between the crucibles for WO$_3$ precursors and sapphire substrate before and after acid cleaning.

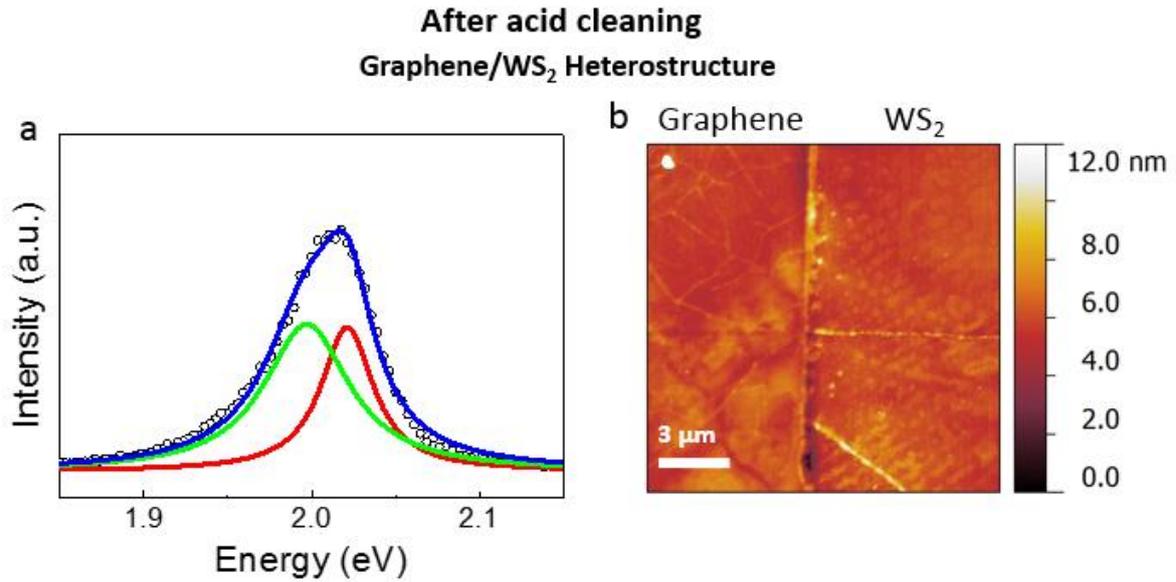

**Figure S4.** (a) $WS_2$ PL spectrum taken from graphene-$WS_2$ heterostructures. Lorentzian fitting shows that the PL peak consists of a trion peak (green curve) and an exciton peak (red curve). Detailed peak information is presented in Table S2. (b) AFM morphology scan of the interface of graphene and $WS_2$ film of the as-grown heterostructure.

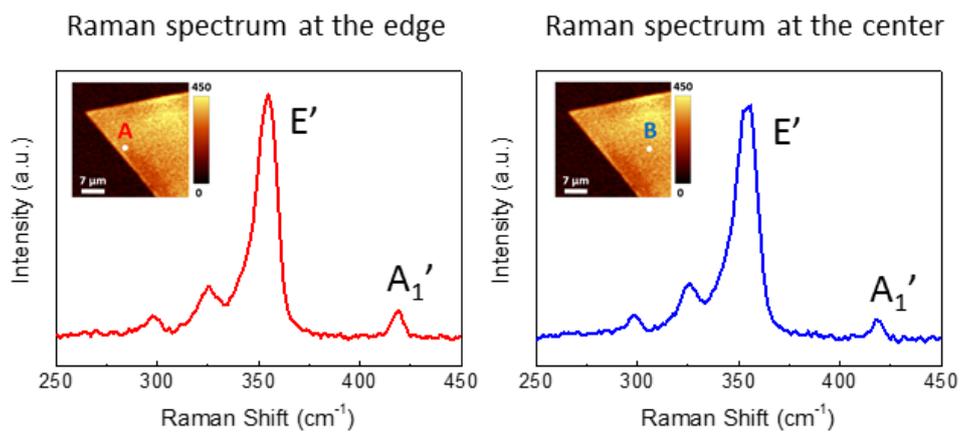

**Figure S5.** Raman spectra taken at (a) the edge (point A, with water intercalation, red curve) and (b) the center (point B, without water intercalation, blue curve) of the single crystal grown by the acid cleaned CVD tube.

Table S1. Spectra information for Figure 4e-f. The PL intensity of the center of the small flake is normalized to 100.

| | | Biexciton | | | Trion | | | Exciton | | | Biexciton binding energy (meV) | Trion binding energy (meV) | Exciton to trion intensity ratio |
| | | Peak position (eV) | Peak intensity (a.u.) | Peak width (meV) | Peak position (eV) | Peak intensity (a.u.) | Peak width (meV) | Peak position (eV) | Peak intensity (a.u.) | Peak width (meV) | | | |
|---|---|---|---|---|---|---|---|---|---|---|---|---|---|
| **Small flake** (Fig 4a-d) | Edge | | | | 1.984 | 575 | 42 | 2.000 | 737 | 31 | | 16 | 1.28 |
| | Center | | | | 1.986 | 76 | 53 | 2.007 | 100 | 41 | | 21 | 1.31 |
| **Large flake new tube** (Fig 1b-e) | Edge | 1.930 | 437 | 93 | 1.972 | 645 | 47 | 1.996 | 555 | 36 | 66 | 24 | 0.86 |
| | Center | | | | 1.977 | 47 | 77 | 2.007 | 65 | 46 | | 30 | 1.39 |
| **Large flake acid-cleaned tube** (Fig 2d-g) | Edge | 1.952 | 649 | 64 | 1.980 | 715 | 38 | 2.000 | 764 | 30 | 48 | 20 | 1.07 |
| | Center | | | | 1.983 | 190 | 51 | 2.008 | 199 | 40 | | 25 | 1.05 |

Table S2. Detailed peak information in Figure S4a.

| | **Peak position (eV)** | **Exciton to trion intensity ratio** | **Exciton to trion width ratio** |
|---|---|---|---|
| Exciton | 2.021 | 0.98 | 0.45 |
| Trion | 1.997 | | |